\documentstyle[12pt,aaspp4]{article}
\def\gsim{\;\rlap{\lower 2.5pt
 \hbox{$\sim$}}\raise 1.5pt\hbox{$>$}\;}
\def\lsim{\;\rlap{\lower 2.5pt
   \hbox{$\sim$}}\raise 1.5pt\hbox{$<$}\;}
\newcommand\beq{\begin{equation}}
\newcommand\eeq{\end{equation}}

\begin{document}

\title{Are HI Supershells the Remnants of $\gamma$-Ray Bursts?}
\author{\bf Abraham Loeb and Rosalba Perna}
\medskip
\affil{Harvard-Smithsonian Center for Astrophysics, 60 Garden Street,
Cambridge, MA 02138}

\begin{abstract}

Gamma-Ray Bursts (GRBs) are thought to originate at cosmological distances
from the most powerful explosions in the Universe.  If GRBs are not beamed
then the distribution of their number as a function of $\gamma$-ray flux
implies that they occur once per $\sim (0.3$--$40)$ million years per
bright galaxy and that they deposit $\ga 10^{53}$~ergs into their
surrounding interstellar medium.  The blast wave generated by a GRB
explosion would be washed out by interstellar turbulence only after tens of
millions of years when it finally slows down to a velocity of $\sim 10~{\rm
km~s^{-1}}$. This rather long lifetime implies that there could be up to
several tens of active GRB remnants in each galaxy at any given time.  For
many years, radio observations have revealed the enigmatic presence of
expanding neutral-hydrogen (HI) supershells of $\sim {\rm kpc}$ radius in
the Milky Way and in other nearby galaxies. The properties of some
supershells cannot be easily explained in terms of conventional sources
such as stellar winds or supernova explosions.  However, the inferred
energy and frequency of the explosions required to produce most of the
observed supershells agree with the above GRB parameters.  More careful
observations and analysis might reveal which fraction of these supershells
are GRB remnants. We show that if this link is established, the data on HI
supershells can be used to constrain the energy output, the rate per
galaxy, the beaming factor, and the environment of GRB sources in the
Universe.

\end{abstract}


\centerline{to appear in {\it The Astrophysical Journal Letters}, 1998}

\section{Introduction}

By now there is substantial evidence that Gamma-Ray Bursts (GRBs) originate
at cosmological distances (Metzger et al. 1997) from unusually powerful
explosions (Kulkarni et al. 1998).  Ignoring beaming, the distribution of
the number of GRBs as a function of $\gamma$-ray flux implies that they
occur once per $\sim (0.3$--$40)$ million years per bright galaxy and that
they deposit $\ga 10^{53}$~ergs into their surrounding interstellar medium
(Wijers et al. 1998).

The detection of afterglow emission extending over months after a GRB event
(Galama et al. 1998) is most naturally interpreted in terms of delayed
synchrotron emission from the relativistic blast wave produced by the GRB
explosion in a surrounding medium (e.g., Meszaros \& Rees 1997; Waxman
1997b). The inferred density of the ambient gas, $\sim 1~{\rm cm^{-3}}$, is
comparable to that of the interstellar medium in spiral galaxies (Waxman
1997b).  In such a medium, the expanding blast wave will weaken
sufficiently to be washed out by interstellar turbulence or by moving
interstellar clouds only after it has slowed down to a velocity of $\sim
10~{\rm km~s^{-1}}$.  For a uniform medium of density $n_{1}~{\rm
cm^{-3}}$, the late phase of the blast wave evolution would result
(Chevalier 1974) in a cool thin HI shell expanding at a velocity $V_{\rm
sh}=4.5~{\rm km~{s^{-1}}} E_{54}^{0.71}n_{1}^{-0.80}R_{\rm kpc}^{-2.2}$,
where $E_{54}$ is the initial energy deposited into the gas in units of
$10^{54}~{\rm ergs}$ and $R_{\rm kpc}$ is the shell radius in kpc.  The
material is expected to decelerate down to $V_{\rm sh}=10~{\rm km~s^{-1}}$
at a radius $R_{\rm kpc}= 0.7~E_{54}^{0.32} n_{1}^{-0.36}$ after a time of
$t=(0.31R/V_{\rm sh})=21~E_{54}^{0.32} n_{1}^{-0.36}$ million years.  Since
this radius is somewhat larger than the scale height of a galactic disk,
the expanding shell might blow a hole in the disk and stretch along the
direction of decreasing density (Kompaneets 1960; Chevalier \& Gardner
1974). Moreover, as the remnant lifetime is comparable to a galactic
rotation period, its shape could be distorted somewhat by the galactic
shear (Tenorio-Tagle \& Bodenheimer 1988).

The rate of GRBs per comoving volume in the Universe can be estimated
from their number-flux distribution after making an assumption about
the redshift evolution of their luminosity function. The simplest
model assumes that GRBs are standard candles and allows only for
evolution of their rate.  If GRBs are further assumed to be a
completely non-evolving population then their rate should be as high
as once per (0.3--1) million years per $L_\star$ galaxy; on the other
hand, if their rate is assumed to be proportional to the cosmic star
formation rate then they should occur once per $\sim 40$ million years
per $L_\star$ galaxy (Wijers et al. 1998).  Since the lifetime of a
GRB remnant is several tens of millions of years, we expect to find
one to several tens of remnants per galaxy.  The inferred energy that
a GRB source deposits into the surrounding gas depends on the assumed
efficiency, $\epsilon$, for converting this energy into the observed
$\gamma$--ray radiation. In the no-evolution model one finds (Wijers
et al. 1998) $E_{54}\sim 0.4\times (\epsilon/1\%)^{-1}$ where the
typical BATSE source redshift is at $\langle z\rangle \sim 0.8$, while
for the stellar evolutionary model (Wijers et al. 1998), $E_{54}\sim
10\times (\epsilon/1\%)^{-1}~{\rm ergs}$ and $\langle z\rangle\sim
2.6$.  If GRB sources are beamed and illuminate only a fraction
$f_{\rm b}$ of their sky in $\gamma$--rays, then their rate should be
higher by a factor $f_{\rm b}^{-1}$ and their energy output should be
lower by a factor $f_{\rm b}$.  Existing observational data is
degenerate to the value of $f_{\rm b}$ [although the degeneracy could
be removed by searching for optically-selected afterglows with no GRB
counterparts (Rhoads 1997)].  A direct count of the number of GRB remnants
per galaxy in the local Universe could break this degeneracy and
determine the values of $f_{\rm b}$ and $\epsilon$ from the production
rate and explosion energy of these remnants.  Since almost all GRBs
show X-ray afterglow emission which signals the existence of an
external medium (while optical emission could sometimes be suppressed
due to dust obscuration), interstellar GRB remnants are likely to form
in a dominant fraction of all GRB events.

\section{HI Supershells}

For several decades, 21 cm surveys of spiral galaxies have revealed the
puzzling existence of expanding giant HI supershells (e.g., Tenorio-Tagle
\& Bodenheimer 1988). These nearly spherical structures are 
deficient of interstellar matter in their interiors and have high HI
density at their boundaries which expand at velocities of several tens of
${\rm km~{s}^{-1}}$.  The radii of these shells are much larger than those
of ordinary supernova remnants and often exceed $\sim 1$ kpc.  By
classification (Heiles 1979), they possess inferred kinetic energies of
$\ga 3\times 10^{52}$ ergs, and their estimated ages are in the range of
$10^6$--$10^8$ years.  The Milky Way galaxy contains probably several tens
of supershells (Heiles 1979; Heiles, Reach, \& Koo 1996), and in one case
the estimated kinetic energy is as high as $\sim 10^{54}$ ergs.  A similar
network of supershells, which clearly result from localized depositions of
energy into the interstellar medium, is observed in other nearby galaxies.
The observed number of supershells is consistent with a production rate of
about one per a few million years per galaxy.  All these characteristics
are close to the expected properties of GRB remnants.  Nevertheless, the
existing literature regards the energy source of these supershells as an
unsettled issue and explores other mechanisms for their production -- all
of which have problems in some specific examples [only Blinnikov
\& Postnov (1998) briefly mention the possibility of a connection with GRBs].

Smaller shells of radii $\sim 200$--400 pc and energies $\la 3\times
10^{52}$ ergs are often explained as a consequence of the collective action
of stellar winds and supernova explosions originating from OB star
associations (McCray \& Kafatos 1987; Shull \& Saken 1995).  The activity
in an OB association creates a bubble in the interstellar medium that is
filled with hot gas. The excess pressure inside the bubble gives kinetic
energy to the ambient gas which collects on a shell around the bubble.
After the shell has expanded to a size comparable to the scale height of
the galactic disk, its upper part accelerates and becomes Rayleigh-Taylor
unstable.  Subsequently, a phenomenon called ``blow-out'' might occur
whereby gas escapes from the interior of the bubble into the halo of the
galaxy (Igumentschev, Shuston, \& Tutukov 1990). The shell then fragments
and its surviving filamentary pieces are sometimes referred to as
``galactic worms''.  This stellar model could satisfy the energy
requirements for shells of modest energies.  However, for giant shells with
larger kinetic energies this explanation requires extreme and often
implausible assumptions.  For the Milky-Way supershells, the stringent
constraints placed by both the number of massive stars present in OB
associations and by the detailed structure of the Galactic disk seem to
rule out such associations as the appropriate sources of energy
(Tenorio-Tagle \& Bodenheimer 1988).  In other galaxies that cannot be
imaged at the same level of detail, one could postulate more populous OB
associations and still invoke the same mechanism, but there are known cases
where even the minimal assumptions appear very extreme. Two notable
examples are the giant HI supershells discovered (Rand \& van der Hulst
1993) in NGC 4631, with radii $R_1=0.9$ kpc and $R_2=1.5$ kpc, and kinetic
energies $E_1=(6$--$10)\times 10^{53}$ ergs and $E_2=(2$--$5)\times
10^{54}$ ergs, respectively. If these shells were to be formed by multiple
stellar winds and supernova explosions, the required OB associations would
need to have $(4$--$10)\times 10^3$ OB stars for shell 1 and
$(1$--$3.5)\times 10^4$ stars for shell 2. But associations with $\ga 10^4$
OB stars would be unprecedented for a galactic disk.  The HII region formed
by an association of $\sim 10^4$ OB stars would have an ${\rm H}_\alpha$
luminosity of $6\times 10^{40}~{\rm erg~s^{-1}}$ and rank as the brightest
among thousands of HII regions in nearby galaxies (Kennicutt, Edgar \&
Hodge 1989).  However, such large associations are not impossible (Williams
\& McKee 1997).

An alternative model suggests that impacts of small companions or high
velocity gas clouds on the galactic disk have formed the observed
shells (Tenorio-Tagle 1981).  However, it is unclear how such
collisions could account for the near-complete ringlike appearance of
the supershell boundaries (Rand \& van der Hulst 1993).  Moreover,
the supershell sizes and kinetic energies imply that the collisions
must have been far more energetic than those thought to occur between
high-velocity clouds and the Galactic disk, where the available
energies are $\la 2\times 10^{52}$ ergs. A search for independent
evidence that massive objects might have passed through the disk of
NGC 4631 has failed (Rand \& van der Hulst 1993).

The giant supershells of NGC 4631 are two examples out of several (Rhode,
Salzer, \& Westpfahl 1997) where a more satisfying explanation for the
nature of the energy source is needed.  We suggest that some supershells
are GRB remnants.  In fact, given the number-flux distribution of GRBs and
the interpretation of GRB afterglows as emission from an interstellar blast
wave, it is {\it unavoidable} to find some active remnants with the
properties of the largest supershells in galaxies. The GRB-produced
supershells might however be a minority population. Unfortunately, the
current data on the statistics and properties of supershells is sketchy and
inconclusive and may not be sufficient for drawing any firm quantitative
conclusions.  In the next section we will attempt to illustrate the
potential implications that might result from better HI data on the
supershell properties and better optical data on the absence of
conventional energy sources to explain them.  For this purpose alone, we
will tentatively adopt rough numbers which were quoted in the literature
(Heiles 1979; Tenorio-Tagle \& Bodenheimer 1988) and assume that most of
the observed supershells are produced by GRBs.

\section{Implications for GRBs}

From the size and velocity of the observed supershells we estimate that a
typical GRB event would have to release $\sim 10^{54}~{\rm ergs}$ (and in
rare cases even ten times more energy), assuming that a few to ten percent
of the initial energy ends up in the final kinetic energy of the supershell
and the rest is radiated away (Chevalier 1974).  The inferred output energy
is close to the rest mass energy of a solar mass object, and might possibly
arise from the collapse of a massive star to a black hole (Woosley 1993;
Paczy\'nsky 1998).  The continuous distribution of explosion energies for
shells of different sizes in the interstellar medium, ranging from the
typical output of a single supernova $\sim 10^{51}~{\rm ergs}$ and up to
$\sim 10^{55}~{\rm ergs}$, would be natural in this interpretation. The
upper end of this energy range is comparable to the spin energy of a
maximally-rotating $\sim 10M_\odot$ black hole that could be tapped
magnetically through outgoing jets and produce the observed GRB events
(Paczy\'nsky 1998).  On the other hand, it would be more difficult to reach
this extreme energy output through the coalescence of a binary neutron-star
system. Detailed theoretical modeling of the observed afterglow emission in
some GRBs (Waxman 1997b) implies that only $\la 10^{53}~{\rm ergs}$ are
being carried by highly relativistic matter in the first few months of the
explosion.  Our energy estimate suggests that ten times more energy might
be outflowing with non-relativistic matter that eventually catches up and
energizes the outgoing blast wave on a longer time scale, after the shock
has decelerated to a sufficiently low velocity.  The enhancement factor
might relate naturally to the value of $f_{\rm b}^{-1}$ if the relativistic
material is beamed.  The process that re-energizes the shock should leave a
clear signature on the late-time evolution of the afterglow lightcurve
(Paczy\'nsky 1998).

Typically, there are of order ten supershells detected per galaxy (Heiles
1979; Thilker, Braun \& Walterbros 1998), and their estimated lifetime is a
few tens of millions of years. The inferred values for the explosion energy
($\propto f_{\rm b}/\epsilon$) and production rate ($\propto f_{\rm
b}^{-1}$) of supershells imply a radiative efficiency of $\epsilon\sim
0.4\%$ and no beaming in the no-evolution model for the GRB population,
while they require $\epsilon\sim 1\%$ and a beaming fraction of $f_{\rm
b}\sim 10\%$ for the evolutionary model where the GRB rate follows the
cosmic star formation history. Both models predict roughly the same
(reasonable) value for $\epsilon$, and rule out strong beaming with $f_{\rm
b}\ll 10\%$ because it implies a much larger population of HI supershells
than observed. This latter constraint is robust since the GRB rate
increases as $f_{\rm b}^{-1}$ while the supershell radius declines only as
$f_{\rm b}^{0.32}$ .  We note that even if the energy is injected
impulsively along a pair of relativistically-beamed jets, the
non-relativistic expansion of the eventual HI supershell would be close to
spherical.

The identification of the centers of the youngest supershells can be
used to reveal the probable location of GRB explosions. This could
test models that associate GRBs with star forming regions (Woosley
1993; Paczy\'nsky 1998), or alternatively rule out models which
relate them to active galactic nuclei or to binaries involving a
neutron star -- which should be kicked away from star forming regions
by the time they coalesce (Tutukov \& Yungelson 1994).  It would
also be natural to search for unusual emission that might be
associated with a potential remnant object inside these supershells,
although any such object is likely to get kick velocities far in
excess of the final supershell velocity and hence exit the supershell
boundary. In general, it should be easier to separate GRB remnants
from supernova remnants during the early (e.g.  adiabatic) phase of
their evolution. As it is highly unlikely that the separate explosions
of $\sim 10^3$ supernovae would be synchronized to within a period
shorter a million years, a young GRB remnant would contain far more
kinetic (or thermal) energy than OB associations might account for.

The rate of energy injection by GRBs per unit volume in the current
Universe can be obtained by multiplying their energy output by their
frequency and by the density of $L_\star$ galaxies, yielding $\sim
(10^{54}~{\rm ergs})\times (2\times 10^{-3}~{\rm Mpc^{-3}})/ (2\times
10^{6}~{\rm yr})= 10^{45}~{\rm ergs~Mpc^{-3}~yr^{-1}}$.  This estimate is
based on the fact that most of the current star formation occurs in massive
$L_\star$ galaxies. The result appears to be close to the energy injection
rate per unit volume of $10^{19-21}$ eV cosmic rays (Waxman 1997a), $(0.45\pm
0.15)\times 10^{45}~{\rm ergs~Mpc^{-3}~yr^{-1}}$.  This coincidence
suggests that GRBs might provide the acceleration sites for the
highest-energy cosmic rays (Waxman 1995) and that the accelerated particles
possibly approach equipartition with the magnetic and thermal energy
densities behind GRB shocks.

The above analysis illustrates the importance of getting better data on the
sub-population of HI supershells which cannot be associated with
conventional energy sources.  We have demonstrated that the detailed
properties of this population could determine the basic parameters of GRBs
and possibly hint as to their most-likely energy source.

\acknowledgements

We thank John Raymond and Eli Waxman for useful discussions.  This work was
supported in part by a NASA HST grant, an ATP grant NAG5-3085, and the
Harvard Milton fund.

\end{document}